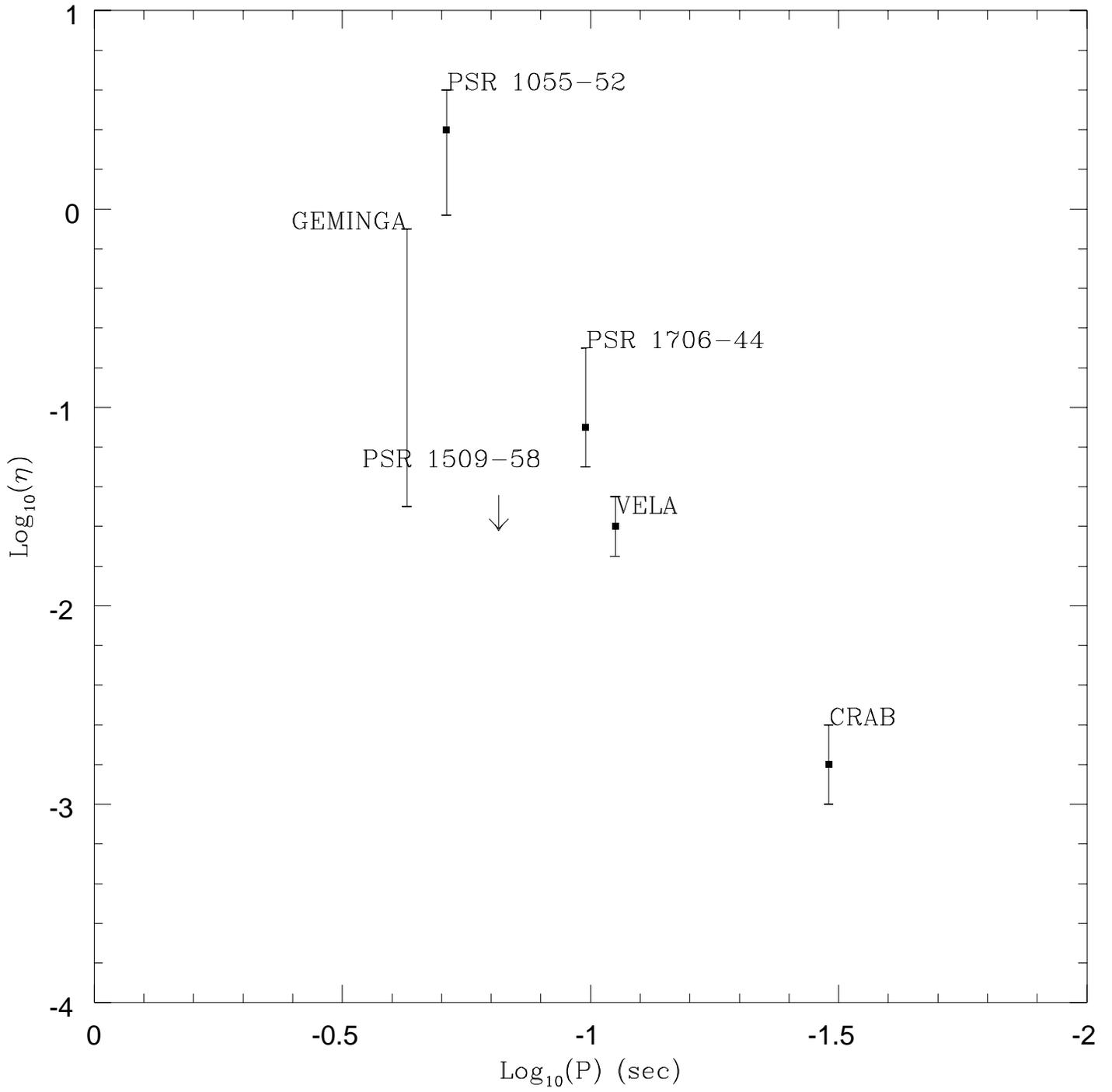

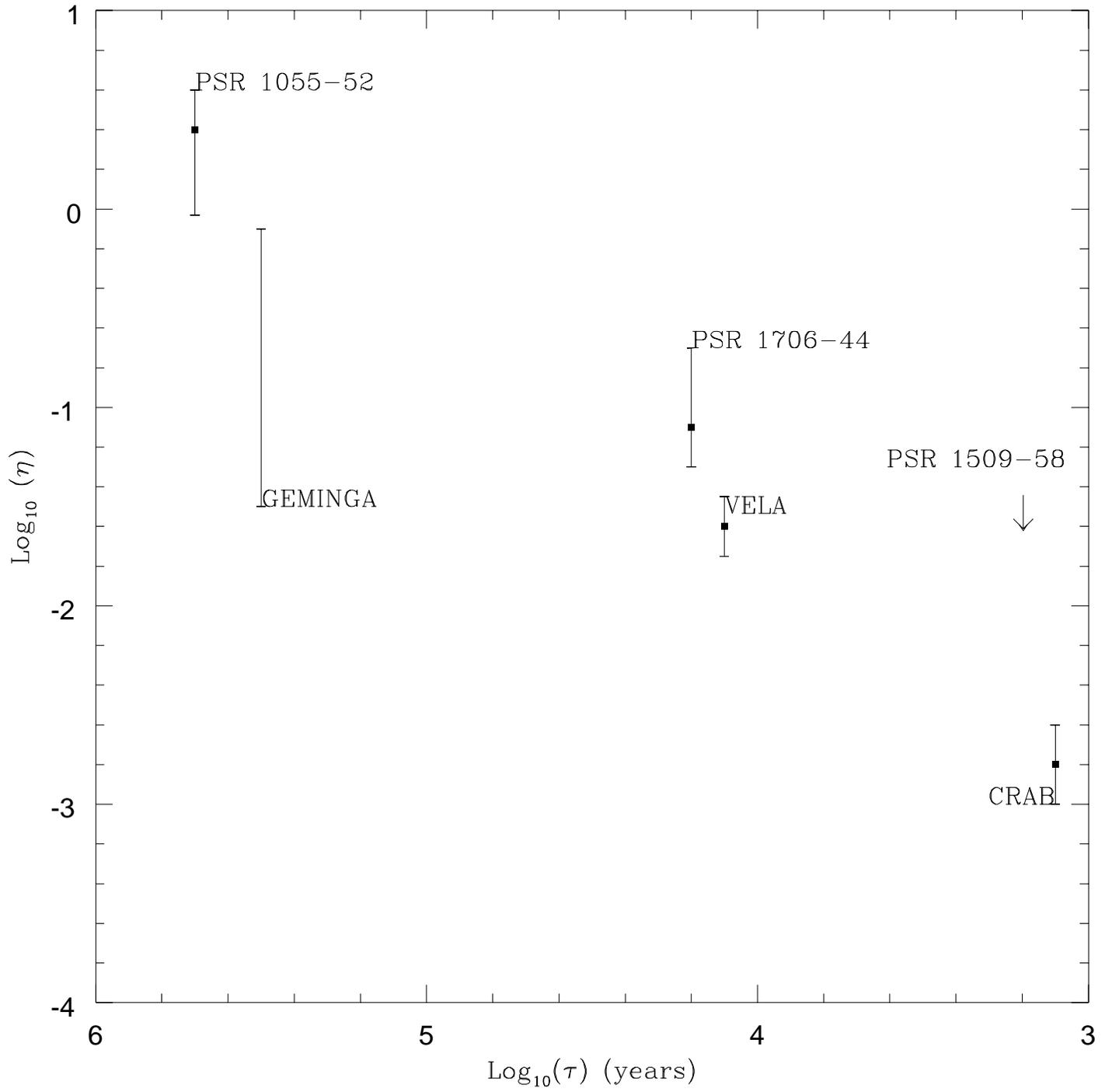

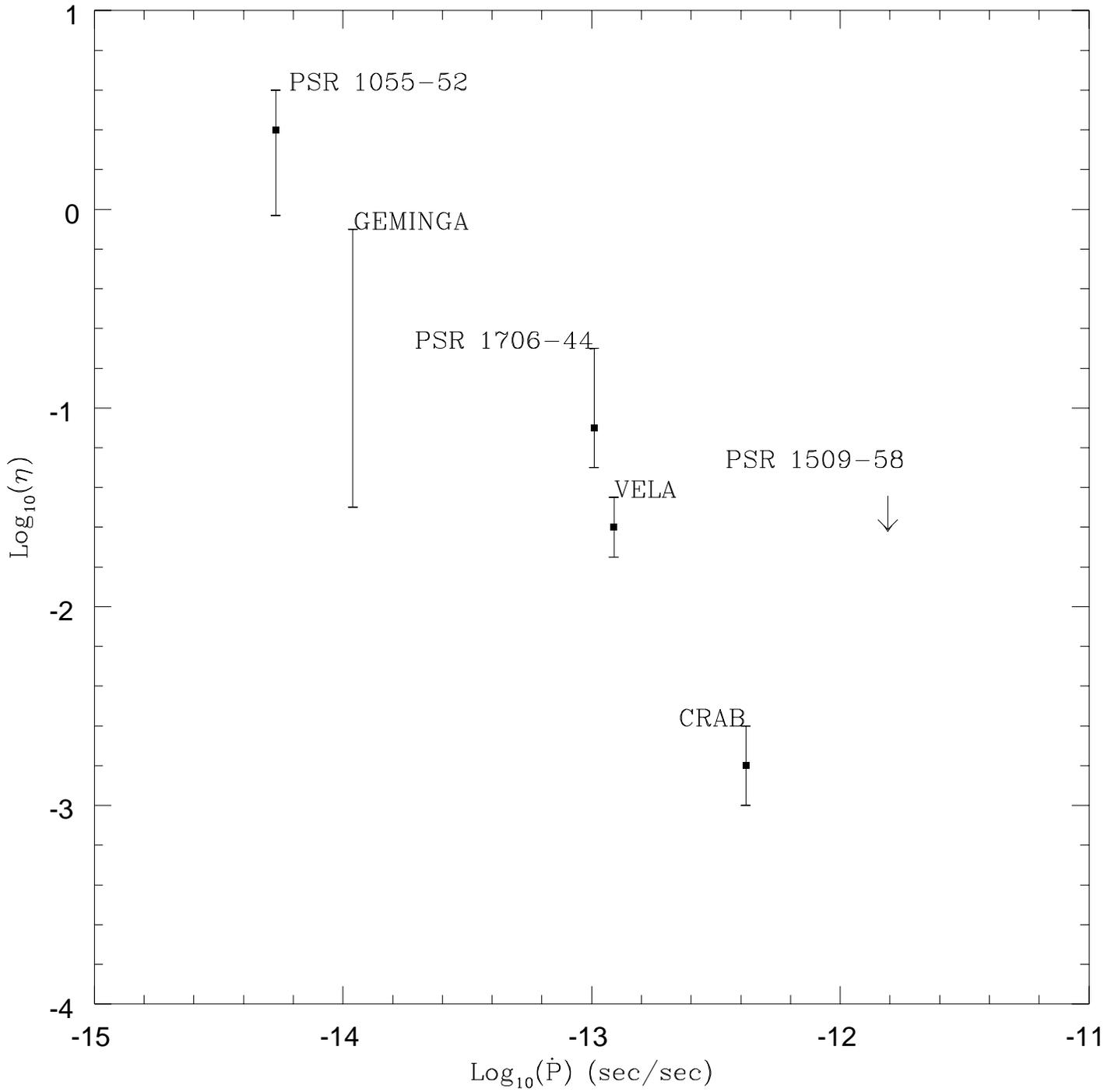

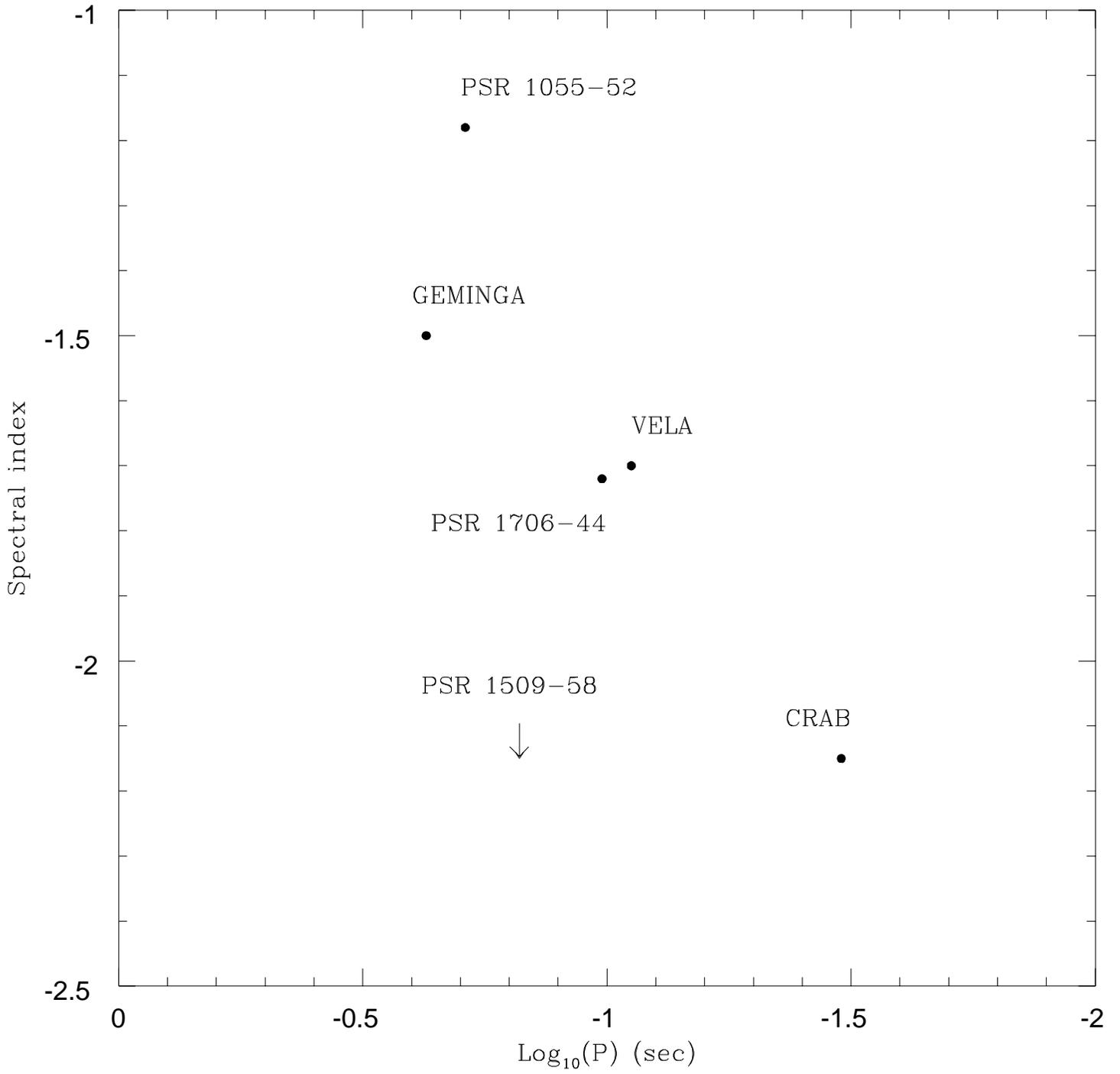

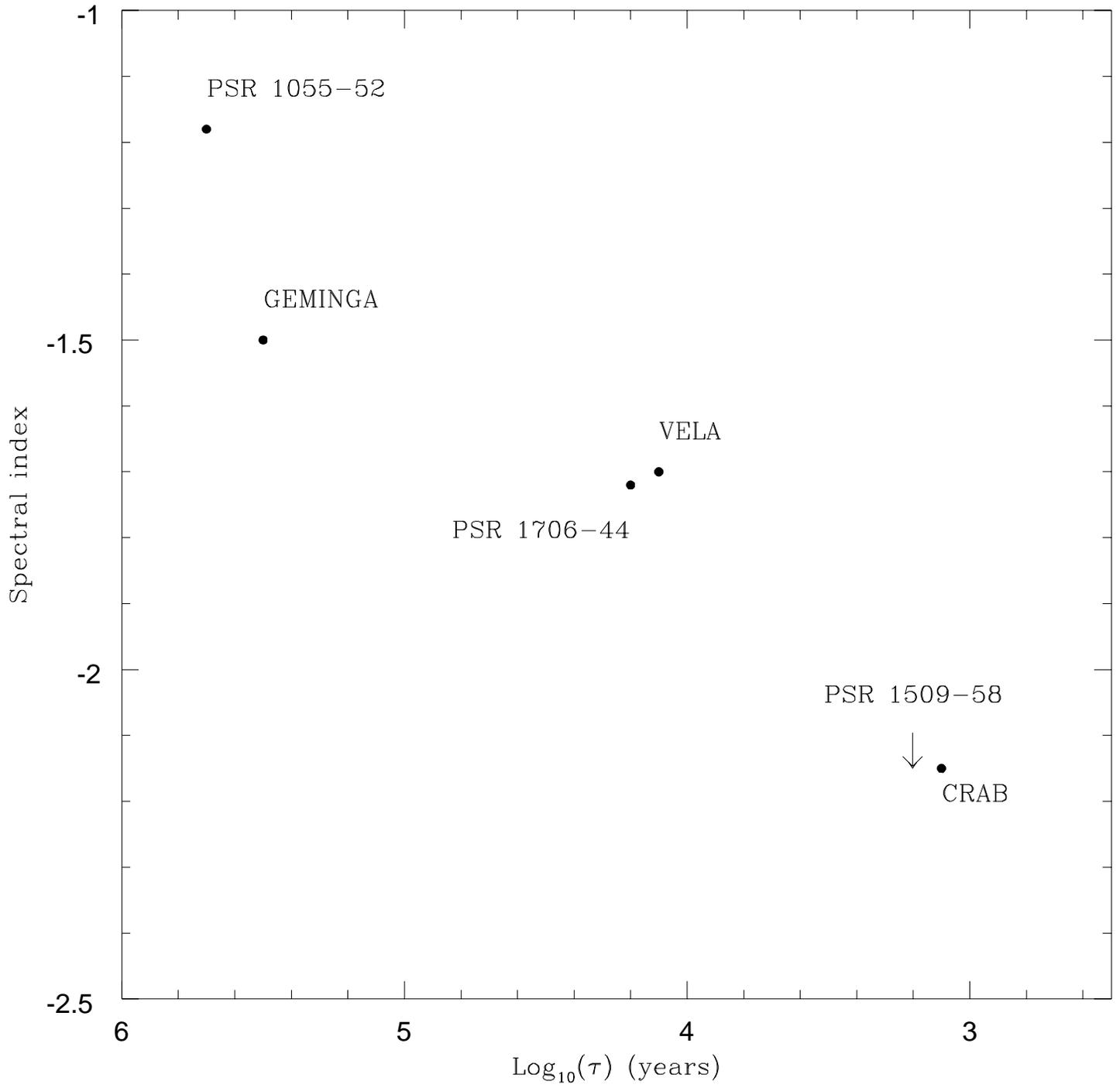

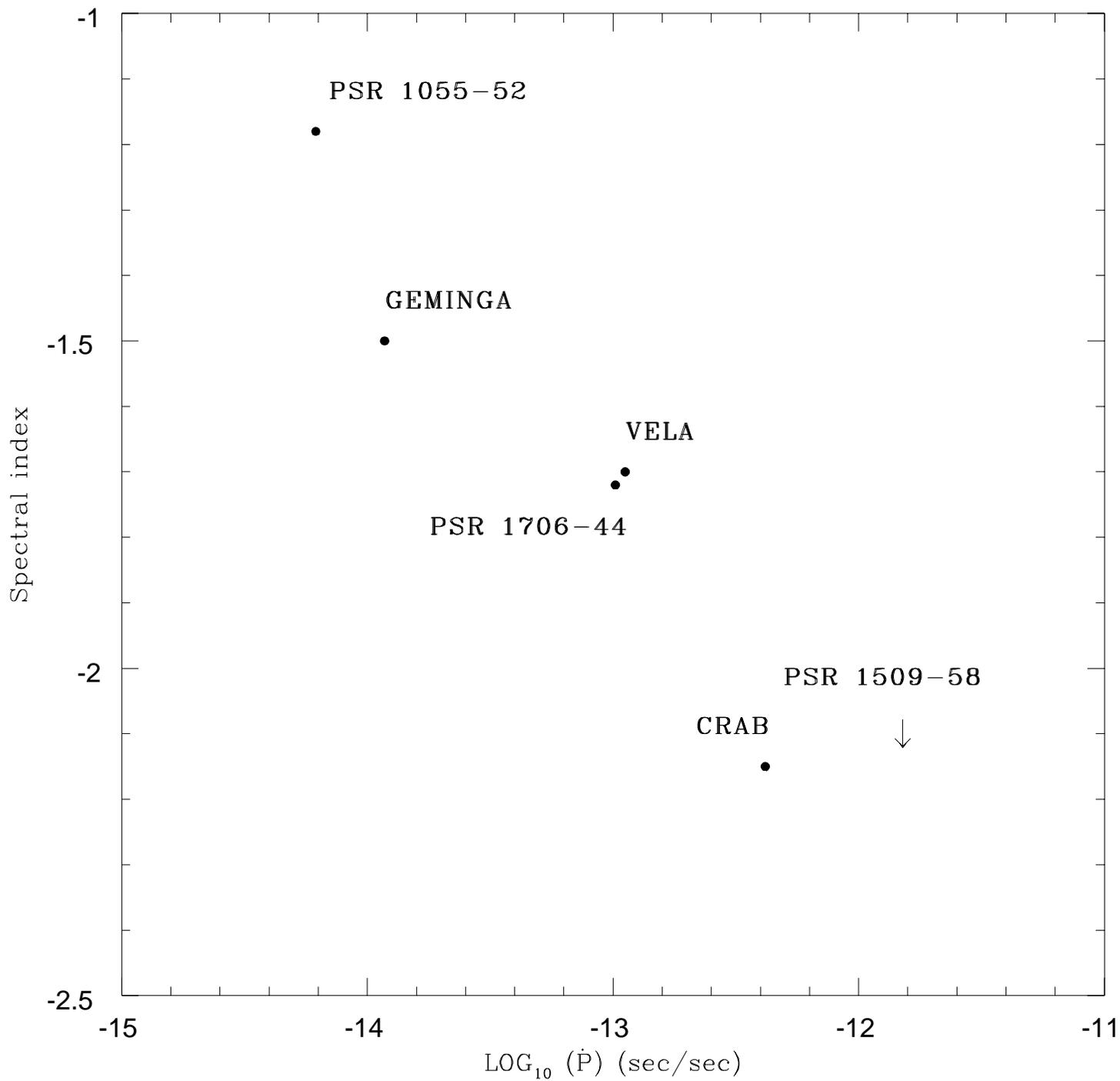

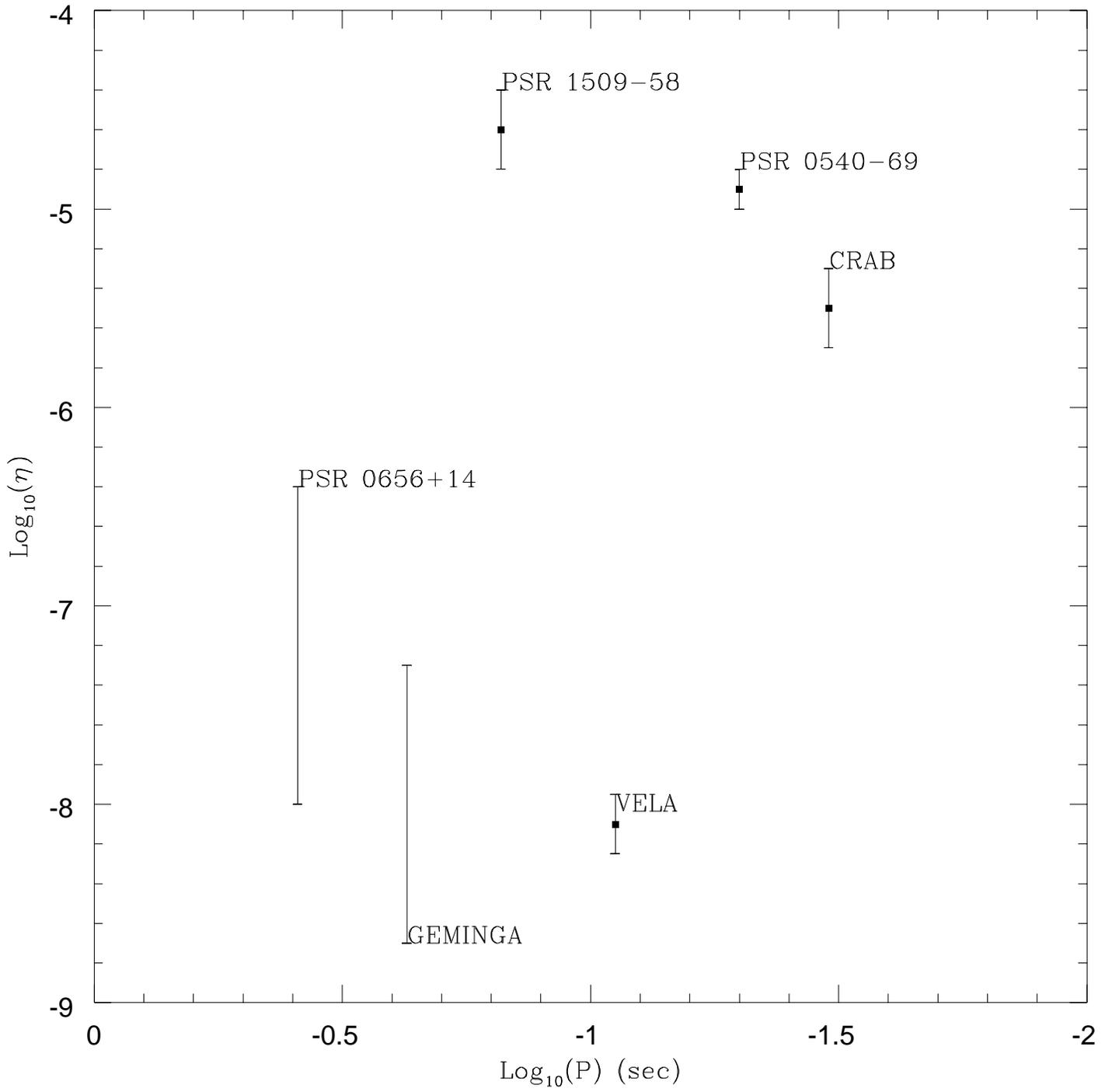

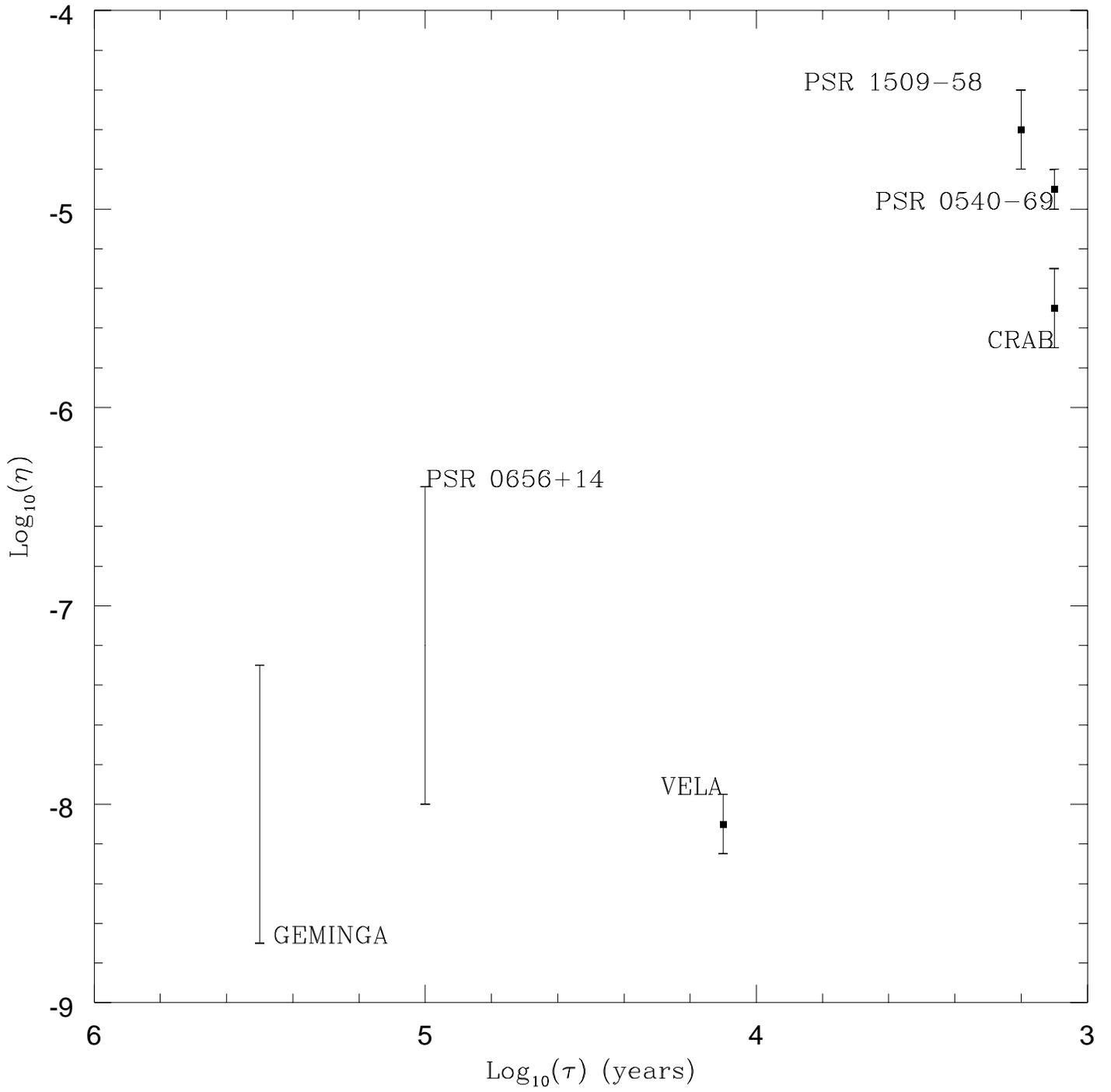

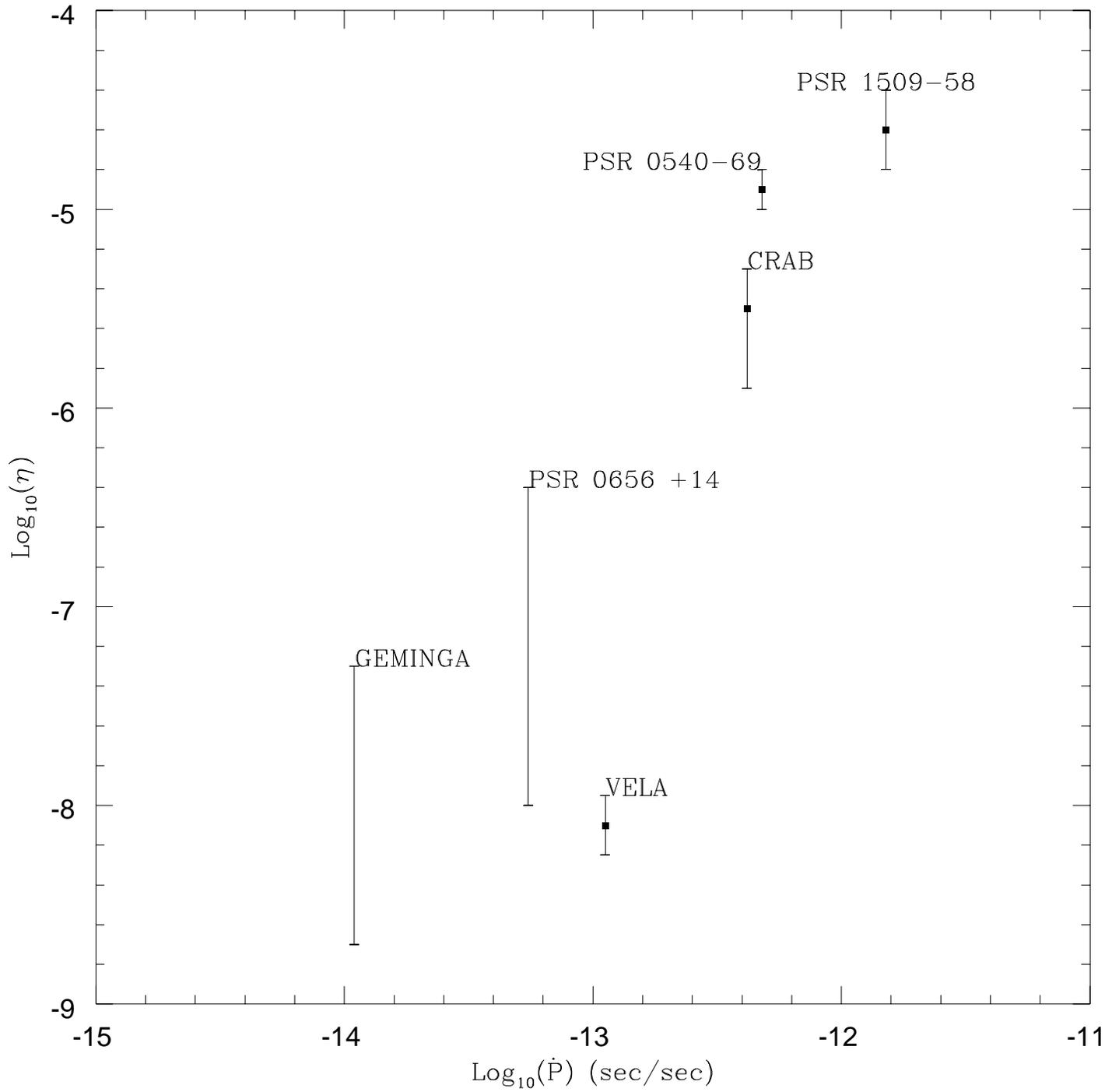

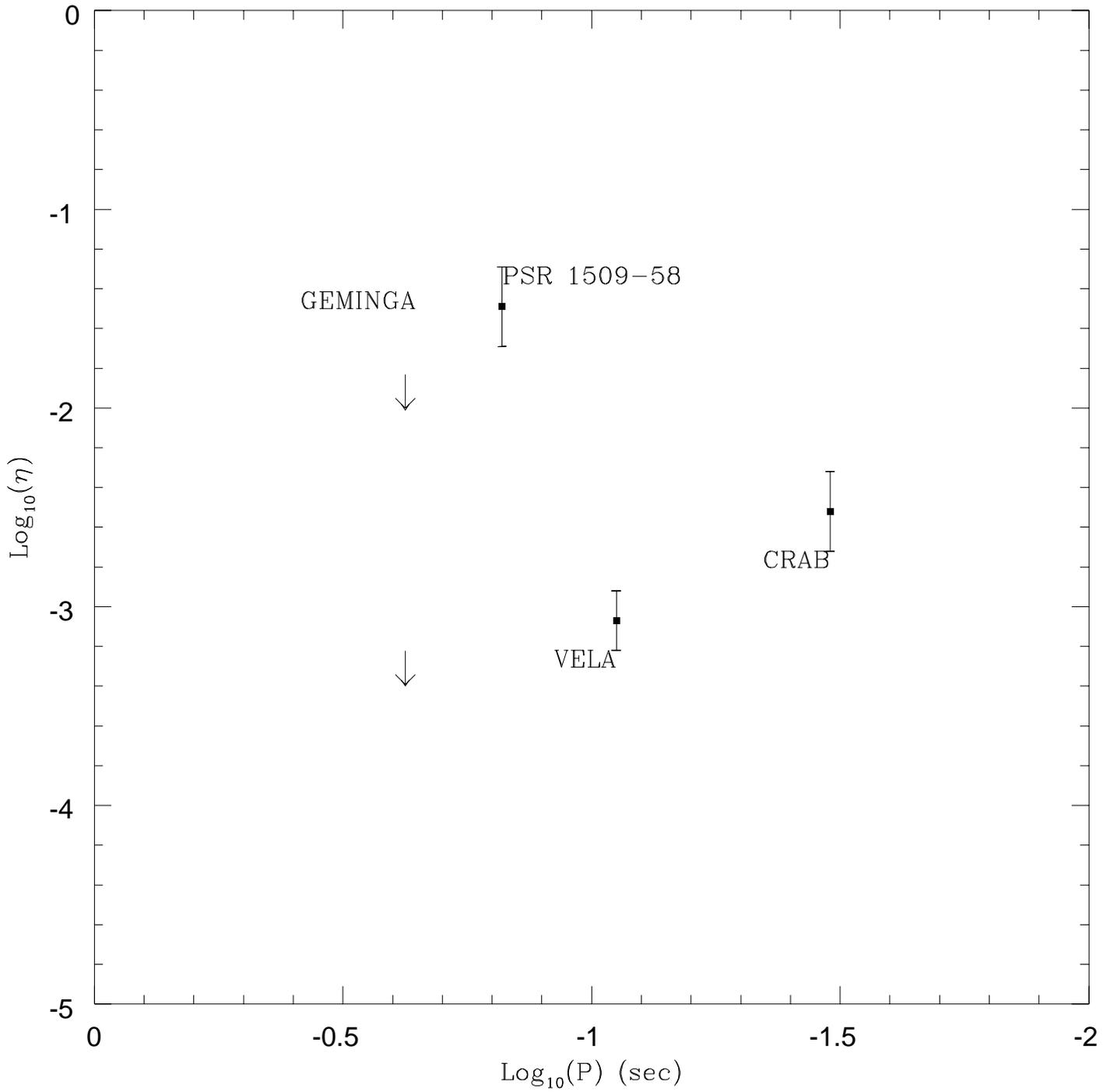

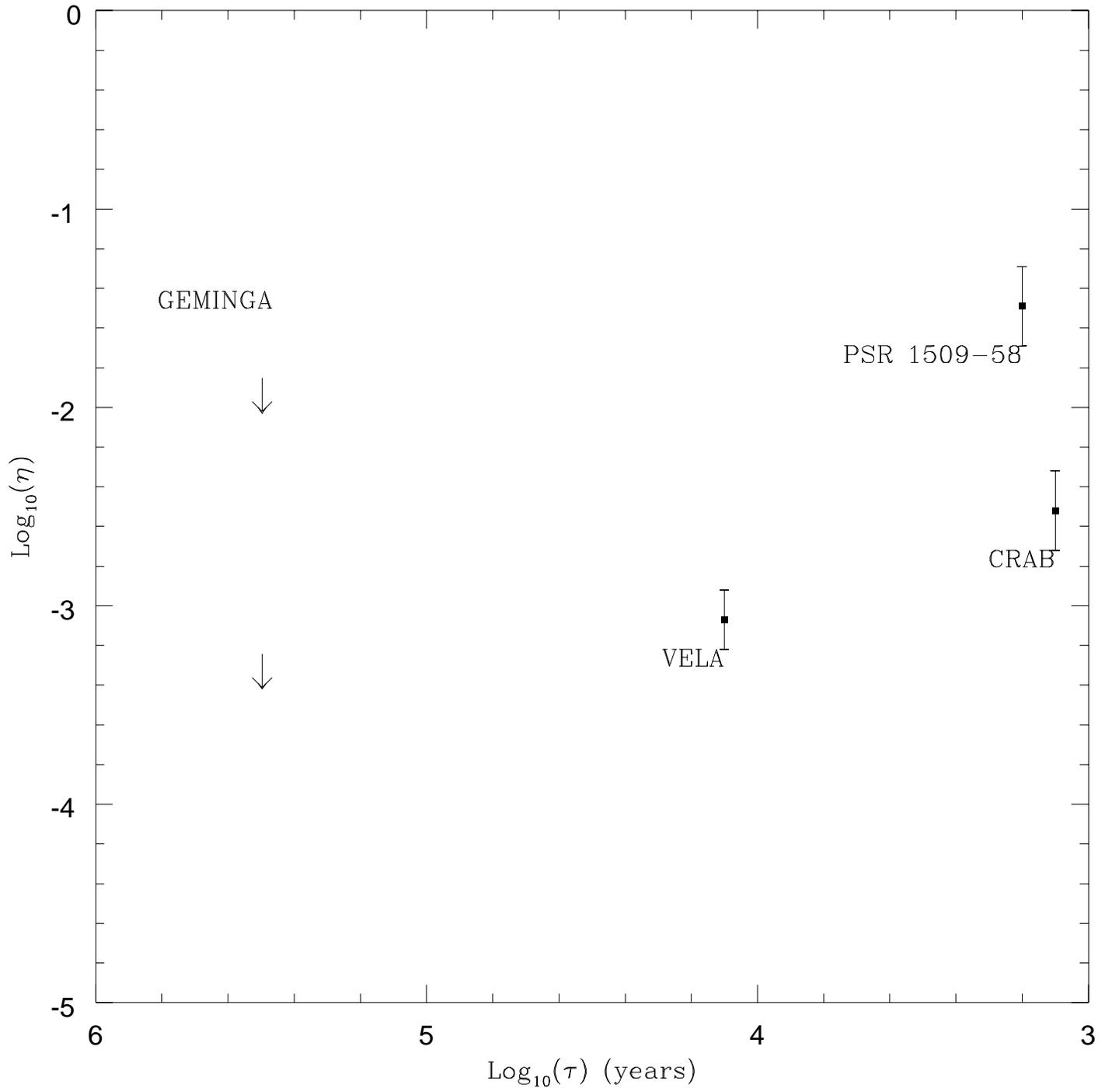

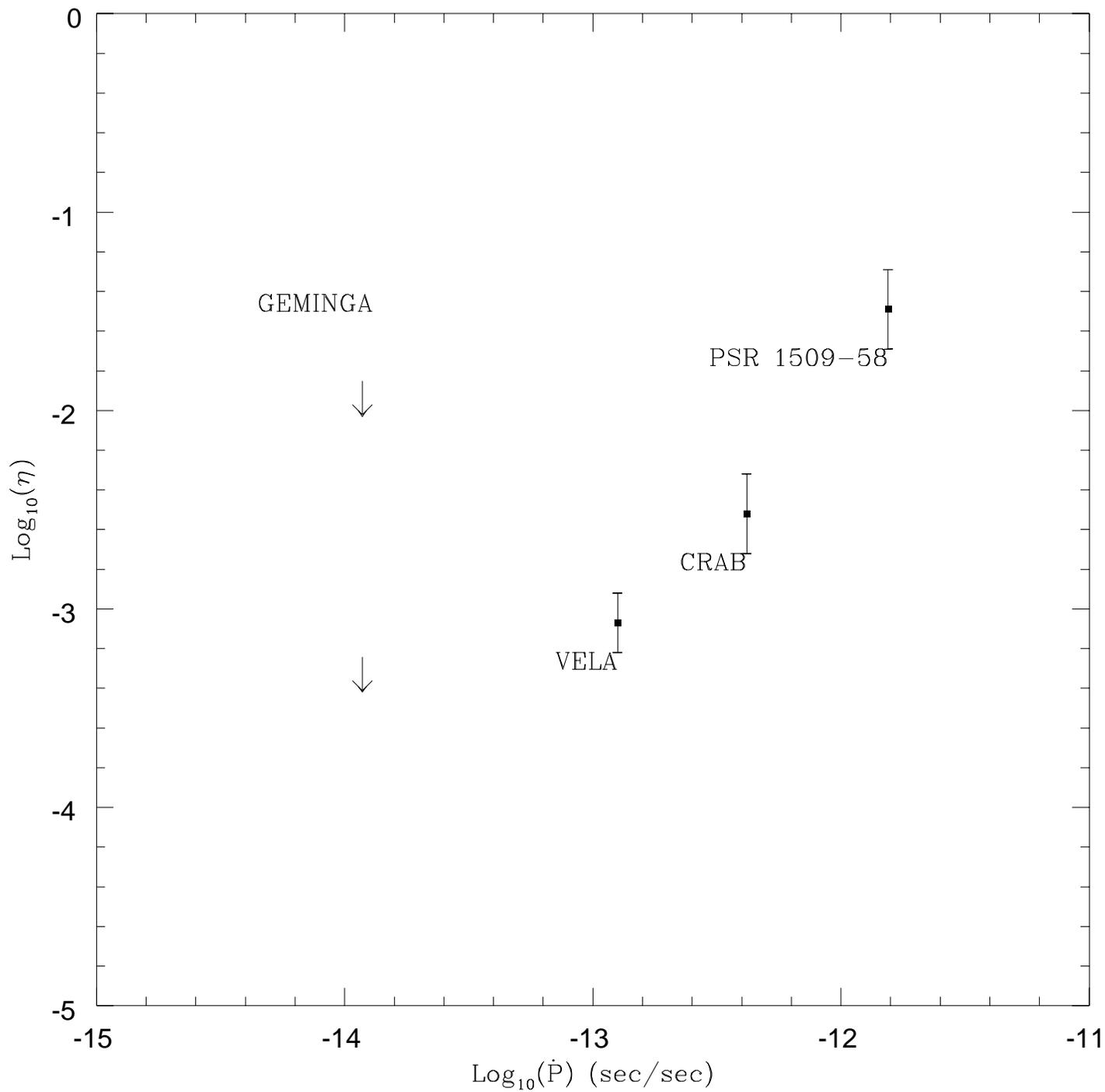

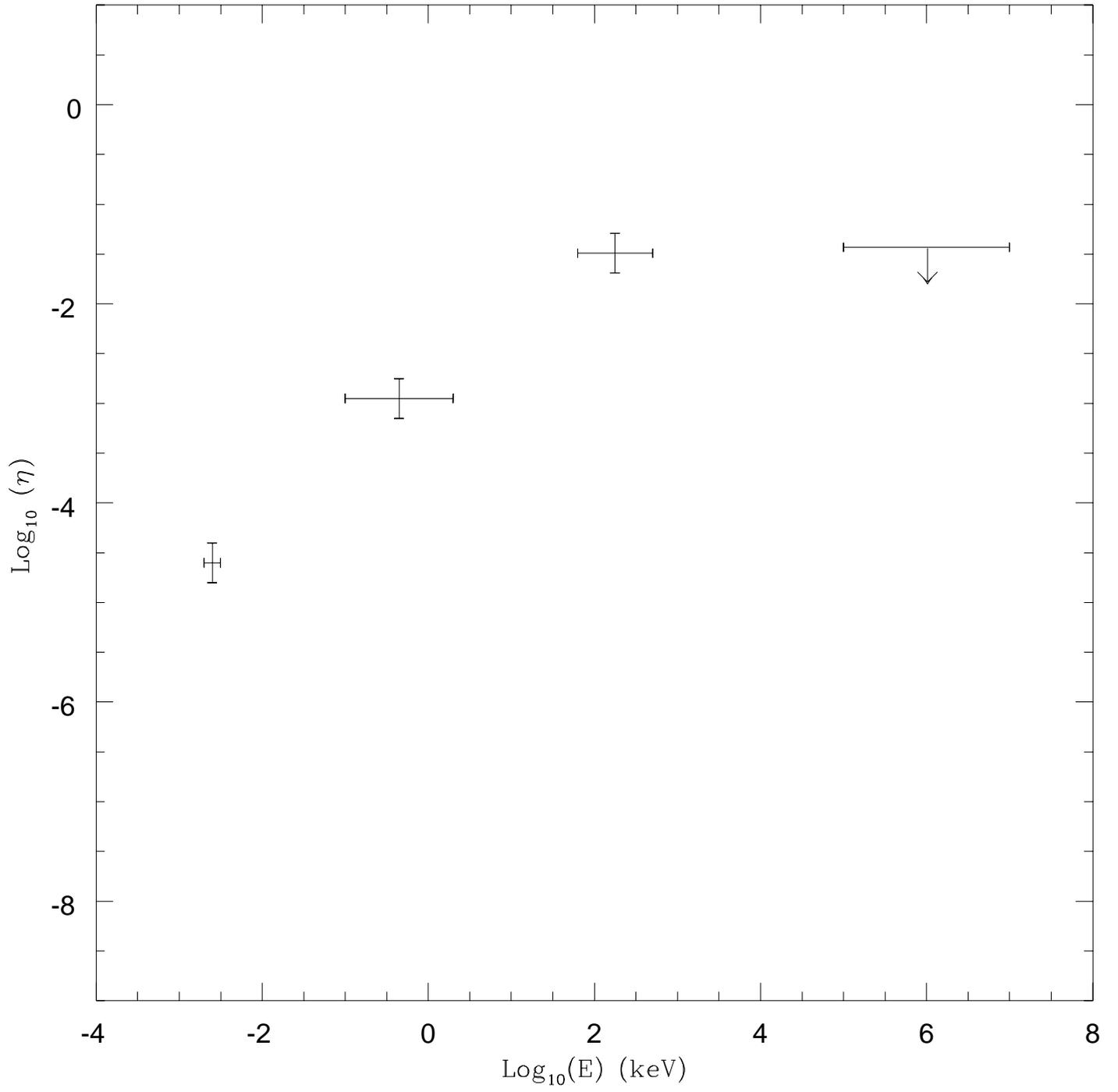

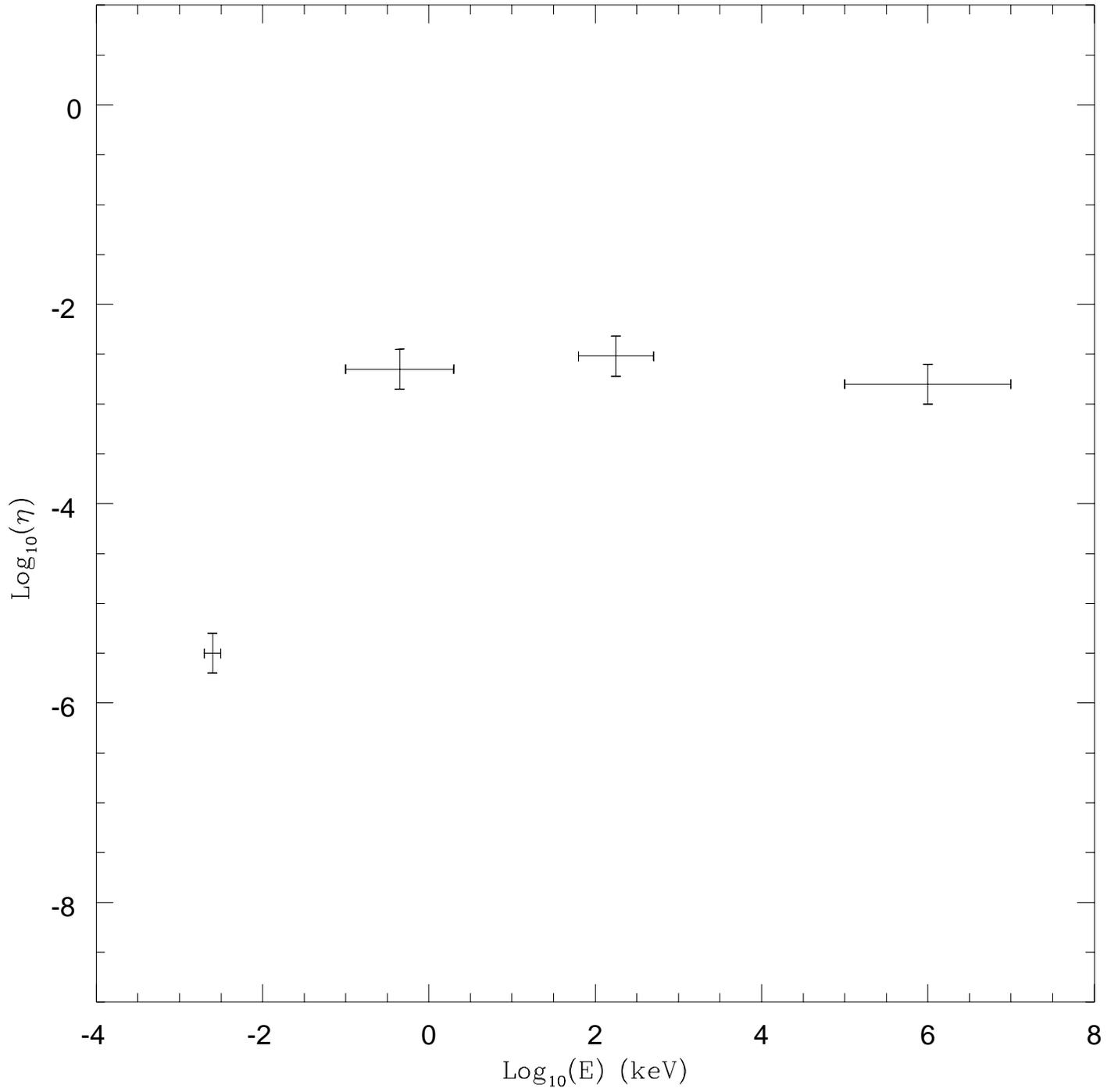
Crab pulsar

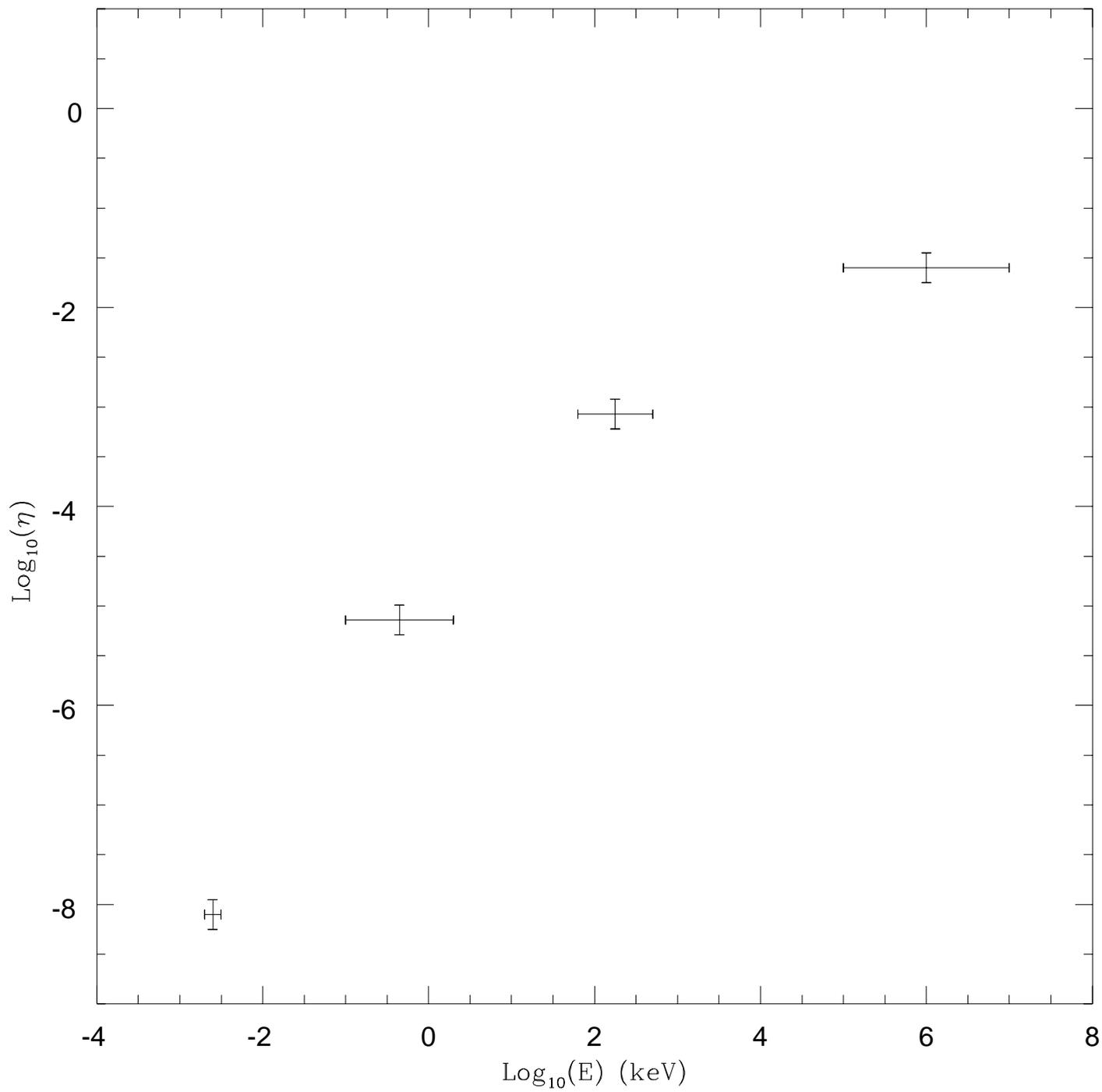

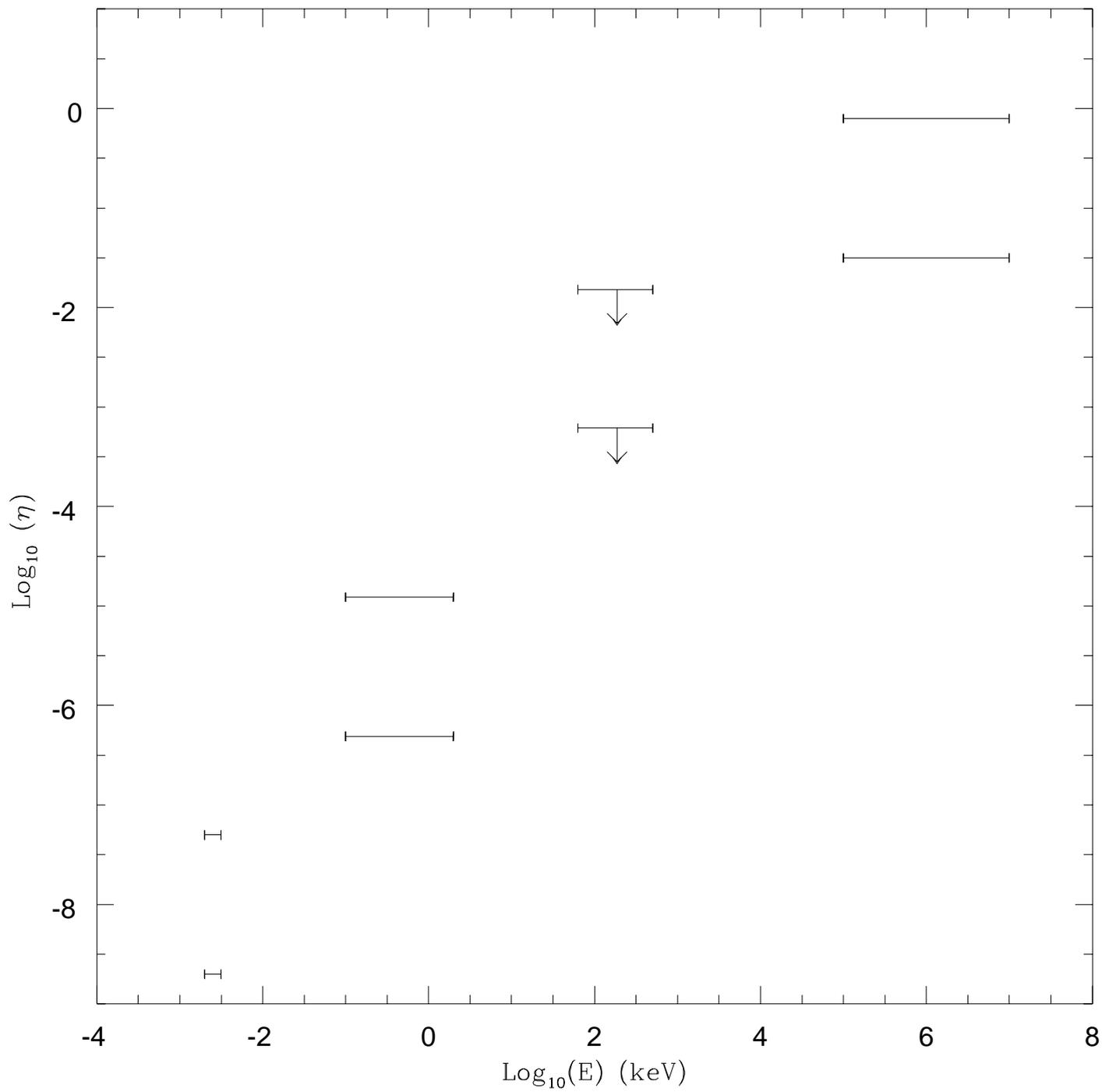

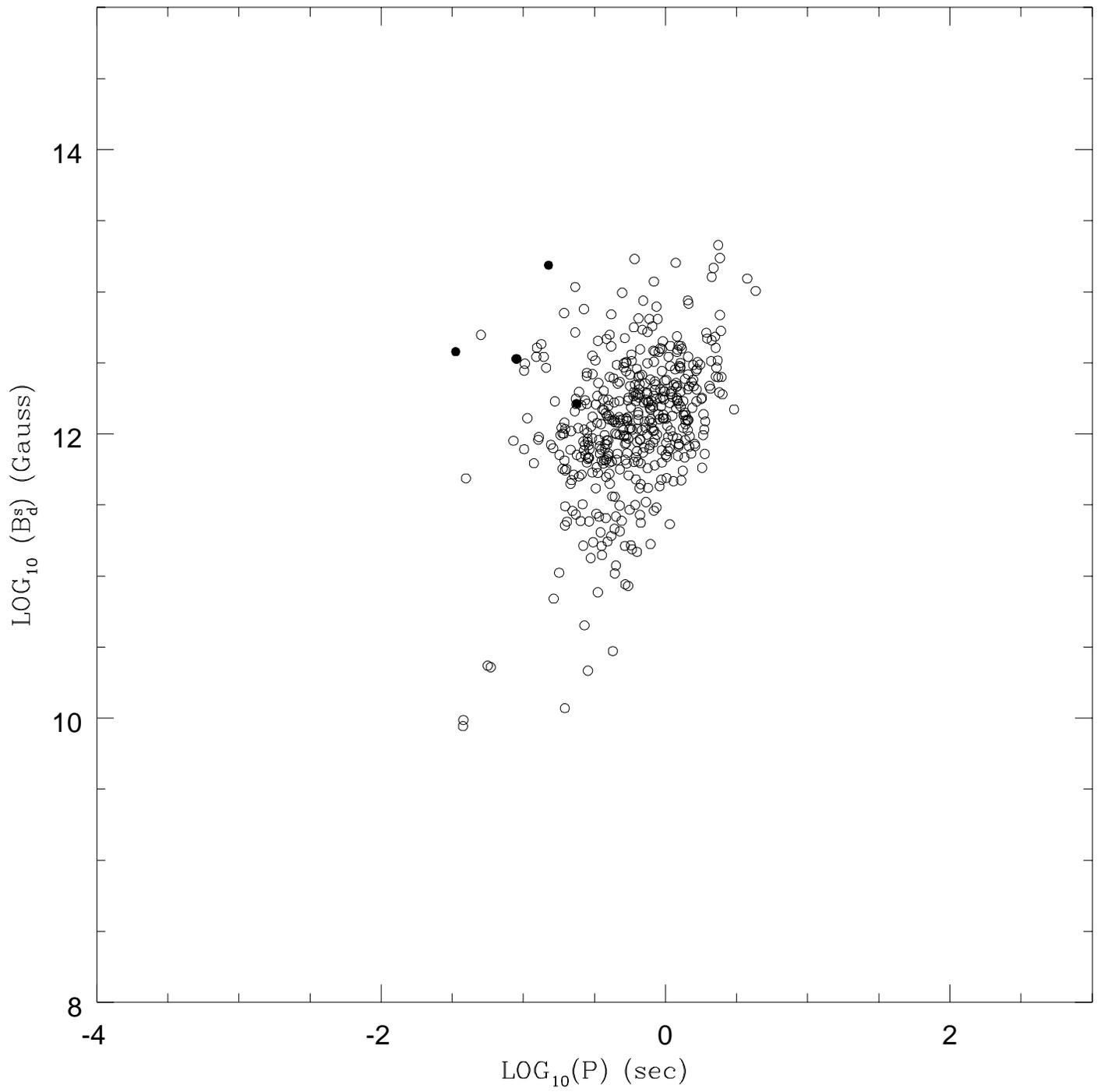

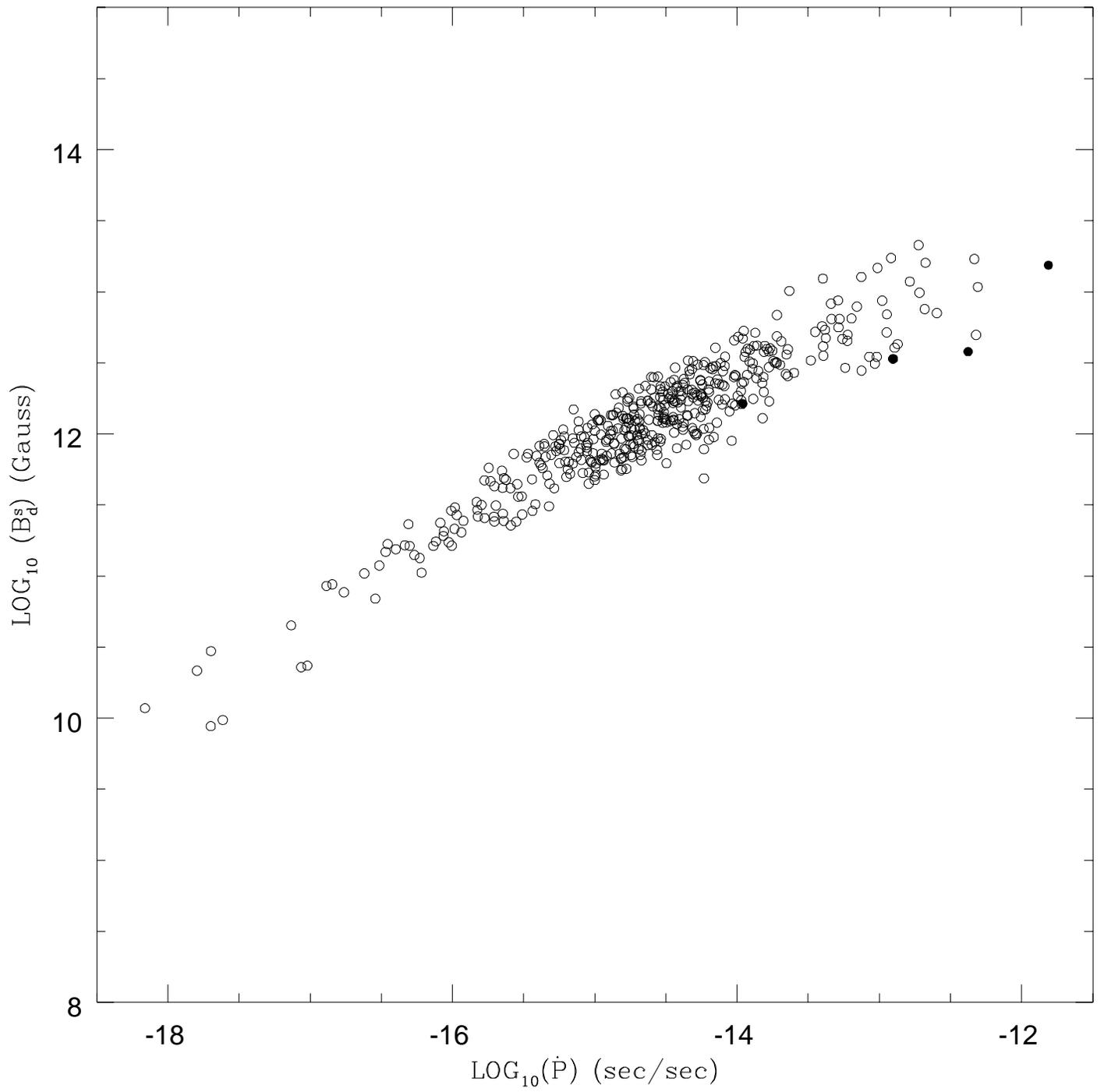

# Multiwavelength Phenomenology of Isolated Neutron Stars

Paolo Goldoni[1,2], Carlo Musso[1,3], Patrizia A. Caraveo[1], Giovanni F. Bignami[1,4]

[1] Istituto di Fisica Cosmica, CNR, v. Bassini 15, I-20133, Milano, Italy; e-mail: (goldoni,musso,pat,gfb)@ifctr.mi.cnr.it
[2] Dipartimento di Fisica, Università di Milano, v. Celoria 16, I-20133, Milano, Italy
[3] Alenia Spazio S.p.A., c.so Marche 41, I-10146, Torino, Italy
[4] Dipartimento di Ingegneria Industriale, Università di Cassino, v. Zamosch 43, I-03043, Cassino, Italy

**Abstract.**
After reviewing the multifrequency behaviour of the Isolated Neutron Stars detected so far, we analyze for each object the efficiency of conversion of the star's rotational energy loss into *optical*, $X$ and $\gamma$ radiation. Although the number of pulsars detected at different wavelengths is rather limited, a pattern is seen to emerge from our analysis pointing towards the period derivative as the leading parameter to describe the multifrequency emission of INS. One object in particular, PSR 1509-58, stands out as archetypal for the $\dot{P}$-dependence of its luminosity at different wavelengths.

## 1. Introduction

Isolated Neutron Stars (INSs) are one of the most exciting topics in astrophysics and have been and continue to be extensively studied in the traditional *radio* domain, the one in which they were originally discovered as radio pulsars.

Recently, the availability of new sensitive detectors allowed the detection of several INSs as *optical*, $X$ and $\gamma$-ray sources. However, in general INSs are faint sources for today's instrument standards and, in spite of the efforts devoted to the subject, only a handful has been detected in the various energy ranges. Of the 558 radio pulsars, 15 have been detected in soft $X$-rays. Of these, 6 have been detected in *optical*, 3 in hard $X$-rays and 5 in high energy $\gamma$-rays. Mereghetti et al. (1994) presents a thorough discussion of the observational data, while Ögelman (1994) gives an updated review of the ROSAT soft $X$-ray data.

Here we shall concentrate on the INSs seen at various frequencies, in order to study in some detail their efficiency of transforming rotational energy loss into electromagnetic radiation. The purpose of the present paper is twofold. Firstly, to understand which INS parameter(s) can best be used to describe their multiwavelength behaviours. It will be apparent that, contrary to common wisdom built over the years, the period derivative and/or the object's age ($\tau = P/2\dot{P}$) are much better than the period itself. Secondly, to point out for the first time, an amazing inversion in the dependence on $\dot{P}$ of the high energy emission from pulsars, taking place in the hundreds of $keV$/ tens of $MeV$ region. Finally, it will be possible to compare the wide-band, multifrequency behaviour of the four best studied INSs, showing very important differences in their *optical*-to-$\gamma$-ray emission distribution. In this context, the important role of PSR 1509-58 will stand out.

## 2. Parameters and observational data for INSs

An INS is a fast rotating object: its rotation period $P$ is therefore important in describing it. Moreover, the star slowly brakes: the time derivative of the period, $\dot{P}$, is another important parameter of the pulsar. On the basis of these two observables, other useful pulsar parameters can be defined: $\tau = P/2\dot{P}$, the "characteristic age", and the magnetic field

$$B_s^d = \left(\frac{3Ic^3}{8\pi^2 r^6}\right)^{1/2} (P\dot{P})^{1/2}, \qquad (1)$$

where $I \sim 10^{45}\ gr\ cm^2$ and $r \sim 10^6\ cm$. Incidentally, we note that the main effect of general relativistic corrections to the magnetic field (Gonthier and Harding 1994) is the increase in its strength by a factor of $\sim 1.5$, without changing its dependence on $P$ and $\dot{P}$.

From an energetic point of view, when we observe an INS, we collect a flux of photons, from which we obtain the isotropic luminosity $L = 4\pi d^2 F$, provided we know the distance $d$ of the source. This can be immediately compared with the rotational energy loss $\dot{E} = I\Omega\dot{\Omega}$ (where $\Omega$ is the angular velocity)$= 4\pi^2 I\dot{P}/P^3$ in order to obtain the efficiency $\eta = L/\dot{E}$, i.e. the capability of the INS of transforming its rotational energy loss into electromagnetic radiation in a given spectral domain.

The sample of INSs seen at different frequencies has grown dramatically in recent years, especially thanks to the sensitive EINSTEIN and ROSAT imaging detectors. In fact, the soft $X$-ray domain is by far the richest in detection of INSs (see Ögelman 1994 for a review). The *optical* follows with 6 detections (see e.g. Caraveo et al. 1994a,b), then we find high energy $\gamma$-rays with 5 detections (see e.g. Thompson et al. 1994) and the hard $X$ rays with 3 (see e.g. Knight et al. 1982, Ulmer et al. 1993, Strickman et al. 1993). Not all the detections are based on timing analysis: half of the identifications in the *optical* domain and about 1/3 of that in the soft $X$-ray domain are based only on positional coincidences, while almost all the hard $X$ and $\gamma$-ray detections rely on pulsation, with the exception of the SIGMA obsevation of PSR 1509-58, which is based on positional coincidence (Laurent et al. 1994).

**Table 1.** Parameters of Isolated Neutron Stars. [1] Thompson et al. 1994; [2] Total efficiency; [3] Upper limit

| Source | P (s) | Log $\dot{P}$ (s/s) | Log $\tau$ (years) | Log B (gauss) | Log $\dot{E}$ (erg/s) | distance (kpc) | Log $\eta_{\rm opt}$ | Log $\eta_{X_{\rm soft}}$ | Log $\eta_{X_{\rm hard}}$ | Log $\eta_\gamma$ | $\gamma - ray$ spectral index[1] |
|---|---|---|---|---|---|---|---|---|---|---|---|
| PSR 1509 − 58 | 0.150 | −11.82 | 3.2 | 13.18 | 37.25 | 4.4 | −4.6[2] | −2.95 | −1.49 | −1.4[3] | −2.10[3] |
| PSR 0540 − 69 | 0.050 | −12.30 | 3.2 | 12.71 | 38.17 | 49.4 | −4.9 | −1.87[2] | | | |
| CRAB | 0.033 | −12.38 | 3.1 | 12.57 | 38.65 | 2 | −5.5 | −2.65 | −2.52 | −2.8 | −2.15 |
| VELA | 0.089 | −12.95 | 4.1 | 12.50 | 36.84 | 0.5 | −8.1 | −5.14 | −3.07 | −1.6 | −1.70 |
| PSR 1706 − 44 | 0.102 | −12.99 | 4.2 | 12.51 | 36.53 | 1.8 | | −4.53[2] | | −1.1 | −1.72 |
| PSR 0656 + 14 | 0.385 | −13.21 | 5.0 | 12.70 | 34.58 | 0.6 | −6.4[2] | −3.53 | | | |
| | | | | | | 0.1 | −8.0[2] | −4.08 | | | |
| GEMINGA | 0.237 | −13.93 | 5.5 | 12.23 | 34.51 | 0.25 | −7.3[2] | −4.91 | −1.82[3] | −0.1 | −1.50 |
| | | | | | | 0.05 | −8.7[2] | −6.31 | −3.21[3] | −1.5 | |
| PSR 1055 − 52 | 0.197 | −14.21 | 5.7 | 12.05 | 34.48 | 1.5 | | −1.62 | | 0.4 | −1.18 |

In the following we shall work with the subset of the soft $X$-ray emitting INSs also detected in $\gamma$-rays, hard $X$-rays or in the *optical* band. This requirement narrows down our source list to eight objects, the parameters of which are given in Table 1. For each object we list $\dot{P}$, $P$, $\tau$, $B$ (no relativistic corrections), $\dot{E}$ together with its distance calculated using the dispersion measurements in the *radio* band (Taylor et al. 1993), with two exceptions. Geminga has not been detected in *radio*: in this case the distance is estimated from the proper motion of the star (Bignami et al. 1993); for PSR 0656+14 the *radio* distance measurement seems to be not consistent with the low column density seen in soft $X$-rays; so we took a range of distances determined by *optical* identification and soft $X$ spectra (Caraveo et al. 1994a). For each object we computed the efficiency in four energy bands, namely: *optical* ($\sim 1\ eV$, V-band), $X_{soft}$ (0.1− to a few $keV$, i.e. the ROSAT/EINSTEIN band), $X_{hard}$ (tens to hundreds of $keV$, OSSE, BATSE, COMPTEL, SIGMA) and $\gamma$ (hundreds of $Mev$ to tens of $GeV$, EGRET, COS-B).

Finally, Table 1 lists also the differential photon spectral index as reported by EGRET in the high-energy $\gamma$-ray domain (Thompson et al. 1994). Note that for PSR 1509-58 the upper limit quoted for the $\gamma$-ray spectral index is to be interpreted as stemming for the comparison between the hard $X$-ray measurements and the EGRET flux upper limits. For these calculations we made use, whenever possible, of the pulsed luminosities, except in the cases for which only total luminosities were available (see Table 1). However, except for the Crab and PSR 0540-69, where a bright plerion is present, for the other INSs considered here the time-averaged luminosities are of the same order of magnitude as the pulsed ones.

The reader can refer to Mereghetti et al. (1994) for a complete discussion of the observational panorama on INSs; here we only summarize the data used to compute the efficiency values given in Table 1.

All the *optical* data come from ground-based telescopes. For the Crab and the Vela pulsars we use the observations of Peterson et al. (1978); the identification of PSR 0540-69 was performed by Middleditch and Pennypacker (1985) and positionally confirmed by Caraveo et al. (1992); the identification of Geminga was proposed by Bignami et al. (1987) and by Halpern and Tytler (1988), and confirmed, owing to proper motion observations, by Bignami et al. (1993); finally, for PSR 0656+14 and PSR 1509-58 we refer to Caraveo et al. (1994a,b). Except for the Crab pulsar, these are all faint objects, with $m_V \geq 22$. It is difficult to obtain for them good spectral or temporal information; in fact, pulsation was detected only for Crab, Vela and PSR 0540-69. Thus, to compute the *optical* efficiencies we have used apparent magnitudes corrected for interstellar absorption, when the latter is known with any degree of certainty.

Note that in the *optical* range there may well be at work more than one physical mechanism. In paticular, while, e.g., the Crab and PSR 0540-69 emissions are non-thermal in origin, objects like Geminga and PSR 0656+14 might have a significant thermal component, as may be the case for soft $X$-rays.

For the $X_{soft}$ band we use the observations made with the PSPC onboard ROSAT in the last four years, except for PSR 1509-58, for which no ROSAT data are available: in this case we refer to the EINSTEIN observations (Seward and Harnden 1982). Converting instrumental count rates into energy fluxes is in general a difficult problem, depending as it does on source spectrum, interstellar absorption and instrument response. We rely on the work of Ögelman (1994), who applies a consistent method to all the detected sources.

In the $X_{hard}$ band, the observations of the Crab pulsar were made with the instruments on HEAO-1 (Knight et al. 1982) and, more recently with SIGMA (Natalucci et al. 1991), OSSE (Ulmer 1994) and COMPTEL (Strong et al. 1993); the Vela pulsar was observed with OSSE (Strickman et al. 1992) and COMPTEL (Bennett et al. 1994), while PSR 1509-58 was seen by OSSE, BATSE (Ulmer et al. 1993) and SIGMA (Laurent et al. 1994); for Geminga there exist only a BATSE upper limit, determined by Wilson et al. (1992). The low number of detections in this energy range is due, basically, to the combination of the low value of the interaction cross section between photons and matter and to the poor signal to noise ratio, due to the impossibility of focussing the radiation and to the presence of high local background.

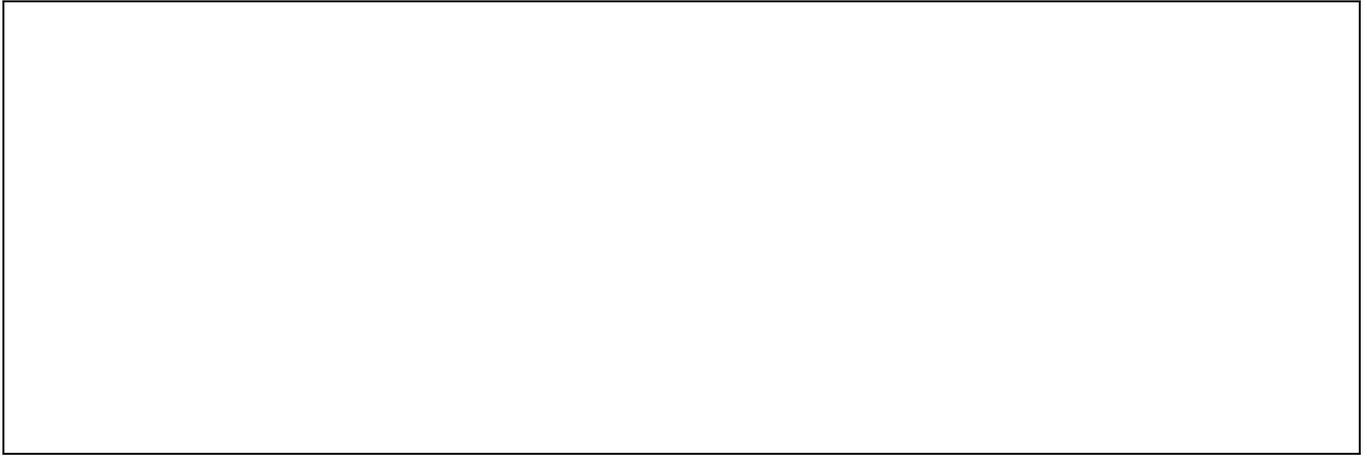

**Fig. 1.** $\gamma$ efficiency ($\eta_\gamma = L_\gamma/\dot{E}$) as a function of $P$, $\tau$ and $\dot{P}$. Error bars reflect uncertainties on the distance values. For Geminga we have plotted $\eta_\gamma$ computed assuming a distance of 50 and 250 $pc$. For PSR 1509-58 an upper limit is shown.

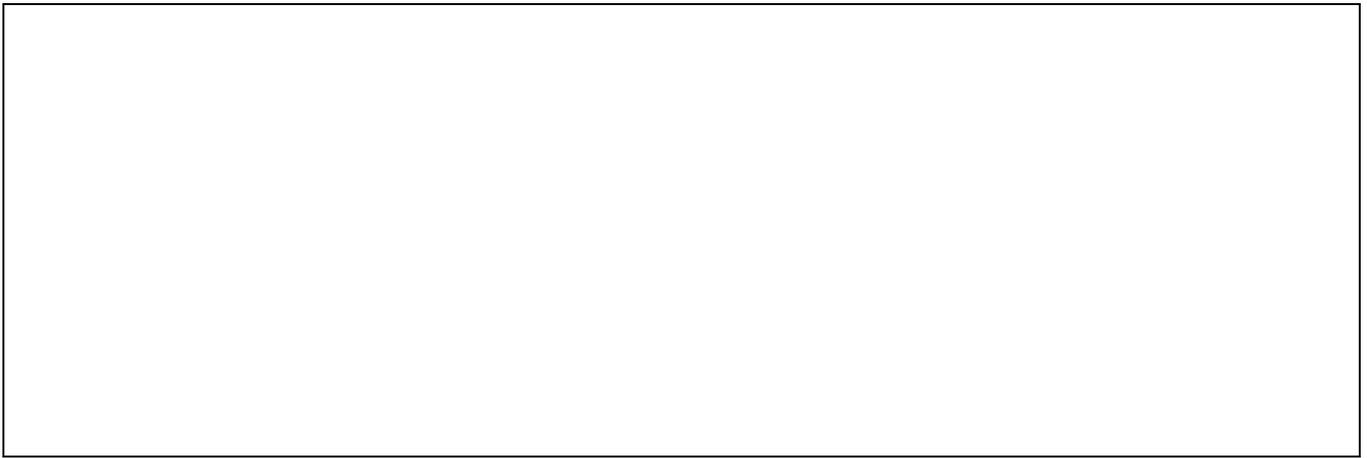

**Fig. 2.** Average $\gamma$-ray spectral indexes of the five INSs detected by EGRET (Thompson et al. 1994) as a function of $P$, $\tau$ and $\dot{P}$

Finally, in the $\gamma$ band the observations come from the COS-B satellite and from the EGRET instrument on the Compton GRO; both cover the range $\sim 100\ MeV-$ tens of $GeV$ albeit with different sensitivities. For the five sources detected in this band, we referred to the following works: Nolan et al. (1993) for the Crab pulsar, Kanbach et al. (1994) for the Vela pulsar, Thompson et al. (1992) for PSR 1706-44, Mayer-Hasselwander et al. (1994) for Geminga, and Fierro et al. (1993) for PSR 1055-52. As noted in Thompson et al. (1994), the increase in $\gamma$-ray efficiencies seem to correspond to an increase in spectral hardness in the $\gamma$-ray domain from tens of $MeV$ to few $GeV$. As mentioned, for PSR 1509-58 only an upper limit exists (Fichtel et al. 1994). For high energy $\gamma$-rays as well as for hard $X$ rays we calculated the fluxes by direct integration of the spectra.

Many of the detections performed are very close to the experimental limits: for example, EGRET collected only 84 photons ($E > 500\ MeV$) from PSR 1055-52 (Fierro et al. 1993), while GEMINGA and PSR 0656+14 have V-magnitudes $\sim 25-26$, i.e. very close to the limit of current ground based (and Hubble Space Telescope) astronomy. Thus, we do not expect a significant growth in the number of identified sources at different wavelengths in the near future: in fact, the first missions likely to truly enlarge the INSs sample (AXAF and XMM for the $X_{soft}$ band, INTEGRAL for the $X_{hard}-\gamma$ band, the VLT for the *optical* band) will be operational after 1999. Therefore, although the number of objects in Table 1 is small, we deem it worthwhile to study this sample.

## 3. Multiwavelength efficiencies

Using Table 1, we search for the best description of the INSs efficiency for transforming $\dot{E}$ into electromagnetic radiation at different wavelengths.

Let us start with the $\gamma$ emission. Fig. 1 gives efficiencies versus the two INS parameter $P$ and $\dot{P}$, and $\tau = \frac{P}{2\dot{P}}$, mainly because of its use in the past pulsar studies. In these plots the error bars on the the $y$-axis are due mainly to the uncertainties on distances (Taylor et al. 1993). We note that the value of the efficiency of PSR 1055-52 is greater than one. This can be ascribed to our neglecting a possible beaming factor. However, according e.g. to Helfand (1994), including an (unknown) beaming factor should not greatly change the situation, so that, for sake of simplicity, we choose not to introduce a free parameter.

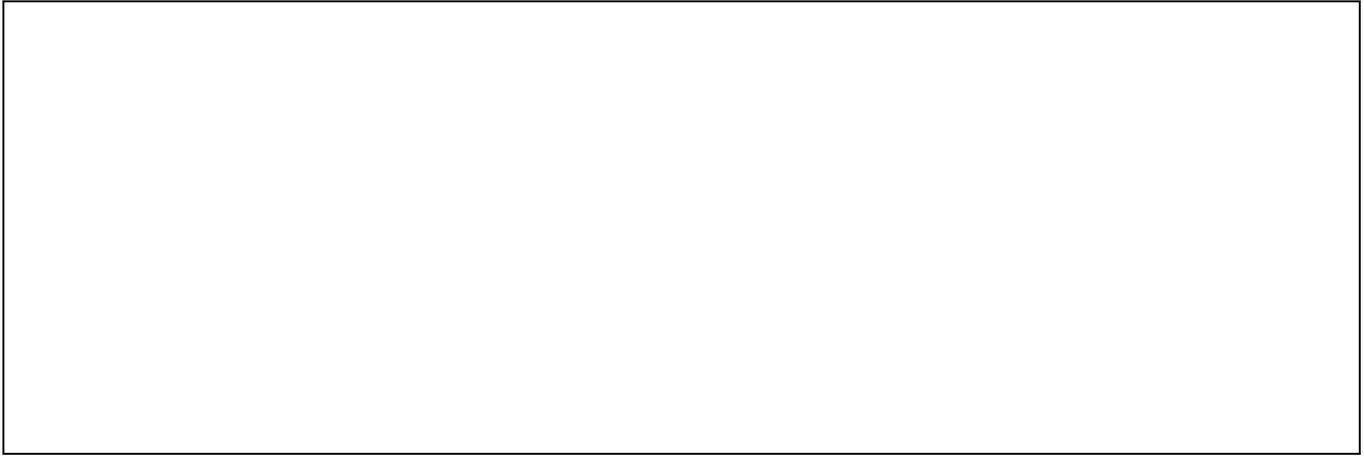

**Fig. 3.** *Optical* efficiency as a function of $P$, $\tau$ and $\dot{P}$. Error bars as in Fig. 1, for Geminga and PSR 0656+14 we have plotted the values corresponding to the maximum and minimum values of the distance, as given in Table 1.

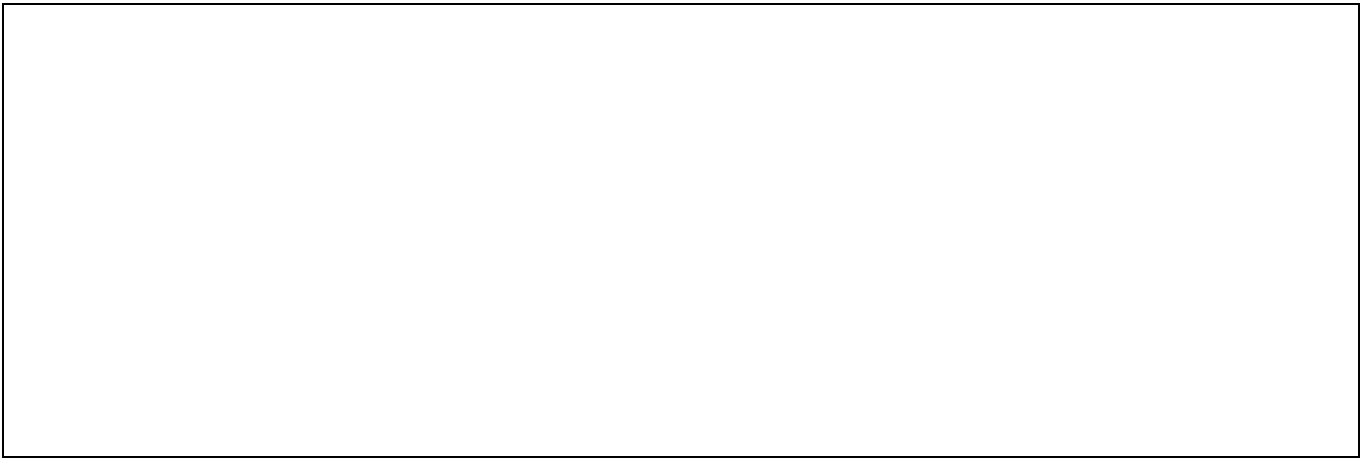

**Fig. 4.** $X_{hard}$ efficiency as a function of $P$, $\tau$ and $\dot{P}$. Error bars as in Fig. 1, for Geminga we have plotted two upper limits corresponding to a distance of 50 and 250 $pc$.

If we ignore, for the moment being, the upper limit for PSR 1509-58, the three plots show a clear trend: $\eta$ increases as $P$ or $\tau$ increase, or $\dot{P}$ decreases, roughly according to power laws, with slopes different in the three cases. The situation somewhat changes if we take into account the upper limit on PSR 1509-58. In fact, the new input does not confirm the trend shown by the $\eta_\gamma$ vs $P$ plot; on the other hand, the high value of this limit does not significantly affect the trend in $\tau$ and $\dot{P}$ plots. This suggests that, in the $\gamma$ band, the right parameters to be used to order efficiencies of different INSs should be $\dot{P}$ (or $\tau$) rather than $P$ (Bignami et al. 1994).

The role of $\dot{P}$ (or $\tau$) is confirmed when observing the $\gamma$-ray spectral index variation, as reported by Thompson et al. (1994) for the five EGRET detections (see Table 1). Fig. 2 shows such a spectral index variation, seen as smoothest as a function of $\dot{P}$. This is especially true if, following Thompson et al. (1994), one interprets the EGRET upper limit on PSR 1509-58 as implying a $\gamma$-ray spectral index (up to 1 $GeV$) steeper than -2.1, as shown in the figure. In fact, PSR 1509-58 appears to have a specially important role also in the *optical* and $X_{hard}$ domains.

Fig. 3 and 4 show the efficiency vs $P$, $\tau$ and $\dot{P}$ for the *optical* and hard $X$-rays bands. Now the trend is in the opposite direction with respect to the $\gamma$-ray case, but the role of $\dot{P}$ shows up much more clearly, largely owing to PSR 1509-58. Both in Fig. 3 and 4 is impossible to find a monotonic behaviour of the efficiency as a function of $P$. In the case of efficiency versus $\tau$ in the hard $X$-ray domain, we would expect a lower value of the efficiency for PSR 1509-58, in order to confirm the trend. As already pointed out by Caraveo et al. (1994b) for the *optical* domain and by Laurent et al. (1994) for the hard $X$ rays one, $\dot{P}$ seems to be better in order to satisfactorily order the *optical* and $X_{hard}$ efficiencies. Nevertheless we consider this evidence not conclusive, especially for the very low number of detections.

In the $X_{soft}$ range the situation is less clear. One must remember that, as mentioned before, in this energy band we may have an unknown mix of thermal and non-thermal emissions, with both pulsed and continuum components (Ögelman 1994). Neither by using raw data, nor by considering only, e.g., the pulsed components, one can obtain a clear trend for $\eta_{X_{soft}}$ as a function of any INSs parameter.

It is interesting to note that PSR 1055-52 has the highest

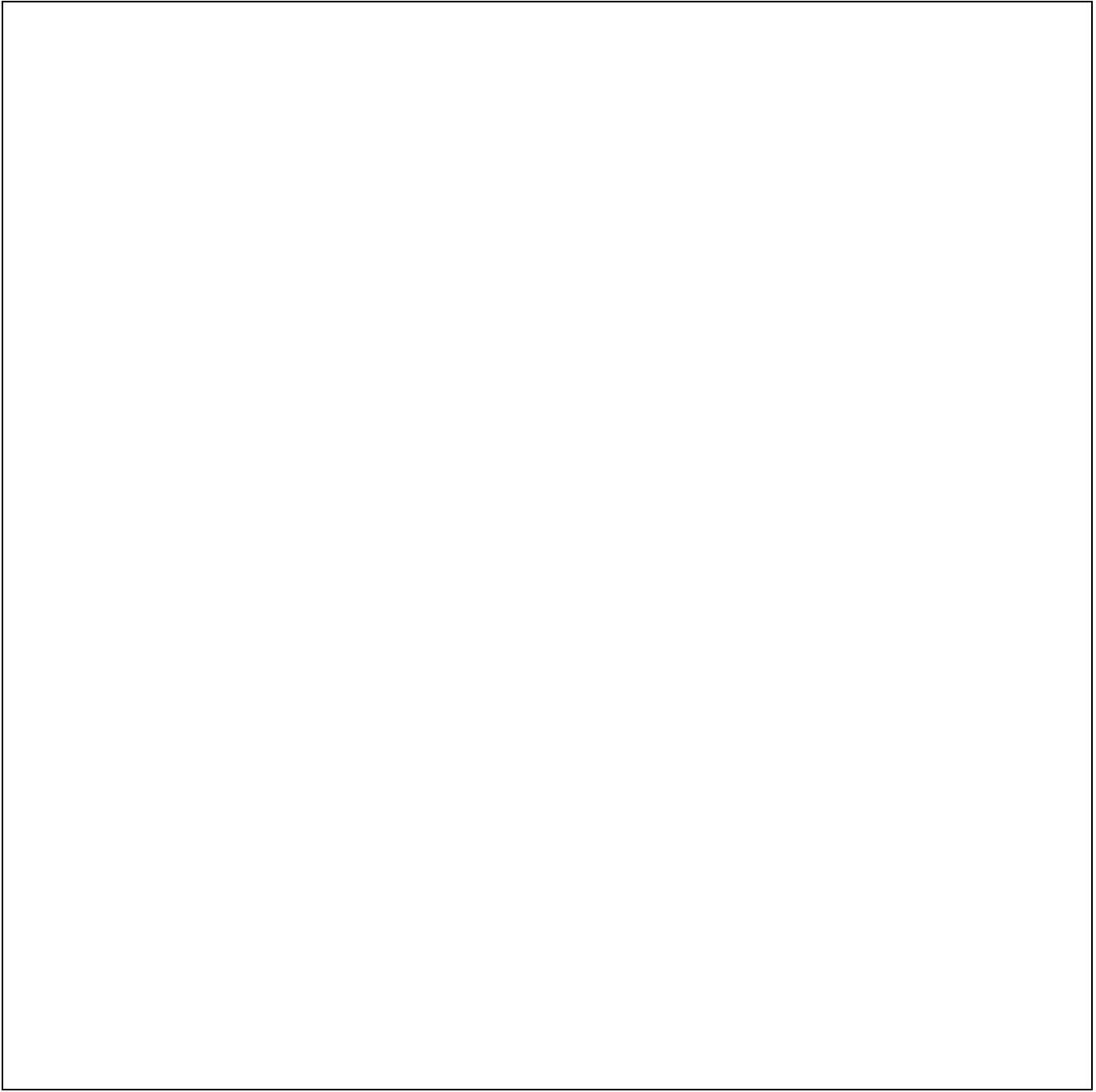

**Fig. 5.** The efficiency of converting the INS rotation energy loss into optical, soft $X$ ray, hard $X$ ray and $\gamma$ ray emission, is plotted vs photon energy for PSR 1509-58, Crab, Vela and Geminga, so far the only INSs detected in at least three of the four energy domains. Horizontal error bars reflect the energy ranges used in Table 1, while the vertical ones are due to distance uncertainties. For Geminga we have plotted the efficiencies corresponding to 50 and 250 $pc$ distances.

$X$-ray efficiency of all, higher even than the Crab pulsar. This might deserve future attention, possibly also in connection with its very high $\gamma$-ray efficiency.

Taking globally the data of Table 1, it appears that the role of $P$ in describing the efficiency of INSs at various wavelengths must be downgraded.

For the few objects for which enough data are available, one can now consider the dependence of efficiencies vs photon energies.

In Fig. 5 we have plotted the efficiency $\eta$ as a function of energy for four INSs, using the same energy bands as defined before. The error bars on the $y$-axis account for distance uncertainties while on the $x$-axis they represent the energy ranges over which we integrated in order to calculate each efficiency value. For Geminga we plotted the efficiencies corresponding to two different values of the distance (50 pc and 250 pc).

The reason for making such plots, instead of the more usual spectral distribution in flux or luminosity, is that in such a way one can more easily see how the rotational energy lost by the pulsar is fractionally distributed along the whole spectrum.

The behaviour of the four objects appears quite different. While the Vela pulsar and Geminga radiate the bulk of their energy in the $\gamma$ part of the spectrum, PSR 1509-58 has a broad maximum in the $X_{hard}$ band and so does the Crab pulsar, showing a kind of plateau from the $X_{soft}$ to the $\gamma$ range. However, looking at these plots in the light of the previous findings, we note that the INSs with higher values of $\dot{P}$ and lower ages, such as 1509-58 and Crab, have the maximum of the efficiency in the $X_{hard}$ domain, while for those with lower $\dot{P}$ and higher ages, such as Vela and Geminga, the maxima are definitely shifted toward the 100 $MeV$ $\gamma$ region. In other words, going from younger and fast decelerating objects to older and slow decelerating ones, there appears an enhancement of the $\gamma$ efficiency and a corresponding suppression of that at lower energies. In spite of the very small number of INSs examined, it is possible to interpret this feature in terms of an evolution of the emission, characterized by the decrease of $\dot{P}$ and the increase of $\tau$. To explain why the efficiency of transforming the rotational energy loss behaves in the way described above is clearly the next challenge for any pulsar emission mechanism.

## 4. Conclusions

Without entering into the details of the different models for electromagnetic emission from a rotating magnetized INS, we look for a simple physical interpretation of the role of our two observables, $\dot{P}$ and $P$, in the emission mechanism.

Let us consider for instance the *polar cap* model (Ruderman and Sutherland 1975; Daugherty and Harding 1982; Arons 1983; Harding et al. 1993). In this case the value of $B_s^d$, which is $\sim 10^{11} - 10^{13}$ $gauss$, greatly influences the emission of an INS, because the particles radiate principally due to the pair production curvature mechanism, which depends on the strength of the magnetic field. It is very reasonable to think that if the magnetic field is strong, the $\gamma$ radiation emitted by the accelerated particles is significantly reprocessed before leaving the star, originating an enhancement of the efficiency in the *optical* and $X$ domains. On the other hand, a lower magnetic field will allow the escape of a less degraded radiation, thus rising the $\gamma$ efficiency.

In order to account quantitatively for the role played by the magnetic field in the pulsar emission process, Lu et al. (1994) have generalized the Hardee (1977) relation into the concept of "generation order parameter". This relates to the number of absorption-reemission cycles a photon must undergo in order to escape the pulsar magnetosphere, and is a simple logarithmic function of $P$ and $\dot{P}$, i.e.

$$\zeta = 1 + \frac{1 - (11/7) \log P + (4/7) \log \dot{P}_{15}}{3.56 - \log P - \log \dot{P}_{15}}. \quad (2)$$

Clearly, then, the higher the $B$ ($\sim \sqrt{P\dot{P}}$) the higher the generation order parameter and the softer the overall radiation escaping from the pulsar. Lu et al. (1994) find a definite correlation between the $\gamma$-ray spectral index and the generation order parameter, which decreases from the value of 3 for the Crab to the value of 1.73 for PSR 1055-52.

Not surprisingly, the highest generation order parameter is found for PSR 1509-58, which appears to be the only pul-sar where as many as 5 cycles of absorption and reemission take place. This naturally explains why PSR 1509-58 is seen up to hundreds of $keV$ but not beyond, and more generally, might also explain the dramatic inversion between hard X-rays and high energy $\gamma$-rays in the $\dot{P}$ dependence, as apparent by comparing Figs. 1 and 4. Of course, such trend inversion is seen reflected in the broad-band energy distributions of Fig. 5. What remains to be understood is the different dependence of the pulsar emissivities on $P$ or on $\dot{P}$.

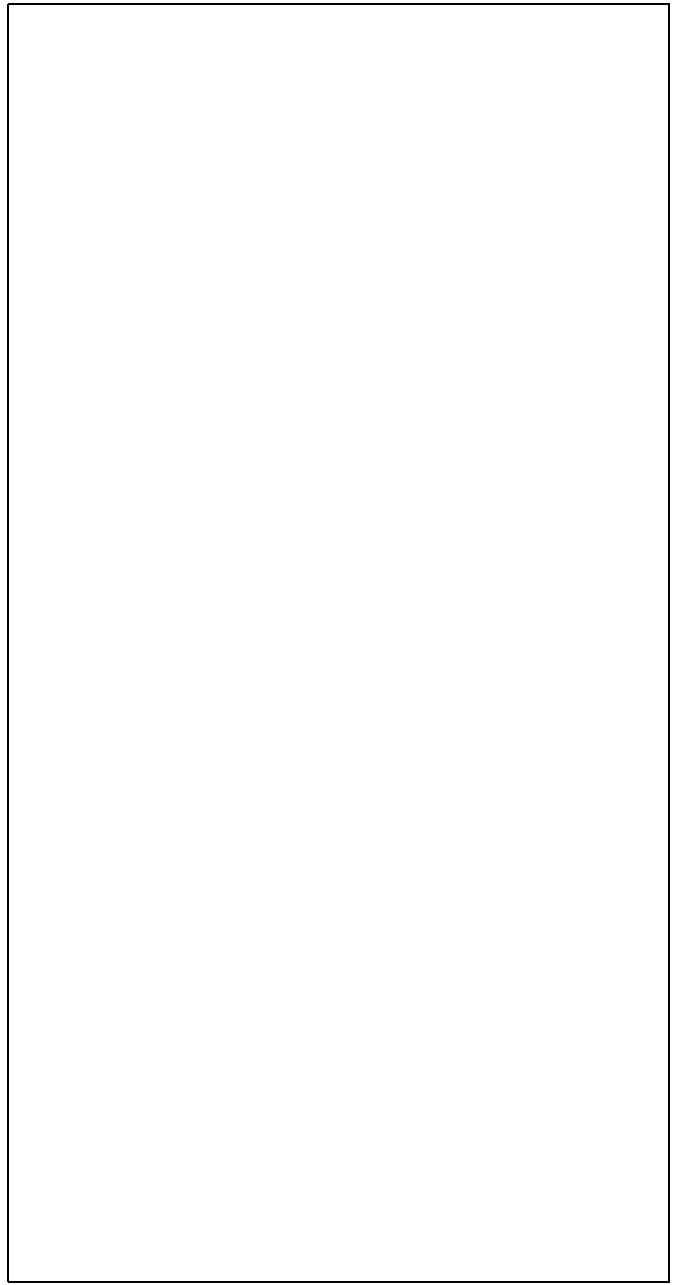

**Fig. 6.** Dependence of the magnetic field $B$ on $P$ and $\dot{P}$, for the 456 (non millisecond) pulsars with known values of $\dot{P}$. The filled circles are the eight INSs listed in Table 1.

In view of the above, it would be natural to directly relate such dependence to the objects' magnetic field. $B$, however,

analytically depends in the same way ($\sim \sqrt{P\dot{P}}$) on both parameters, in contrast with the observed privileged correlation with $\dot{P}$ over $P$.

We thus looked at the $B$ vs $P$ and $B$ vs $\dot{P}$ plots for the population of all (non millisecond) radio pulsars with measured $\dot{P}$, or 456 objects. As seen in Fig. 6, the dispersion in the $B$ vs $P$ plot is far greater than that in the $B$ vs $\dot{P}$ one. Indeed, no correlation is measurable in the first case, while a good fit is possible in the latter, the linear correlation coefficients being $r_P = 0.38$ and $r_{\dot{P}} = 0.93$. This can be understood quantitatively remembering that, for the sample of Fig. 6, the values of $\dot{P}$ span six orders of magnitude, while those of $P$ are restricted to less than two. Thus, $\dot{P}$ will dominate any analytical dependence weighting equally, a priori, $P$ and $\dot{P}$. This is true not only for $B$, but also for the pulsars' age, $\tau = P/2\dot{P}$, also $\dot{P}$-dominated (see comparison of Figs. 1,3 and 4). The conclusion then is that the INS high-energy emission mechanism is dominated by the magnetic field, both in its capacity as photon generator and as photon absorber. PSR 1509-58, with its high $\dot{P}$, may be especially interesting, and worth a lot more studies, for its special role as soft (but certainly not hard) $\gamma$-ray emitter. On the strength of this conclusion, it should also be possible to predict worthwhile pulsar targets for the various instruments aboard the GRO.